\documentstyle[epsfig]{aipproc}

\begin{document}
\title{XMM-Newton Observation of the Black Hole Microquasar GRS 1758-258}
\author{A. Goldwurm$^*$, D. Isra\"el$^*$, P. Goldoni$^*$, P. Ferrando$^*$, A. Decourchelle$^*$, I. F. Mirabel$^*$ and R. S. Warwick$^{\dagger}$ }
\address{$^*$Service d'Astrophysique/DAPNIA, CEA-Saclay, F-91191 Gif-Sur-Yvette, France 
\\
$^{\dagger}$Dept. of Physics $\&$ Astronomy, Leicester University, Leicester, LE1 7RH,  U.K.
}
\maketitle

\begin{abstract}
The XMM-Newton X-ray observatory pointed the galactic black hole 
candidate and microquasar GRS 1758-258 in September 2000 for about 10 ks
during a program devoted to the scan of the Galactic Center regions.
Preliminary results from EPIC MOS camera data are presented here.
The data indicate that the source underwent a state transition from
its standard low-hard state to an intermediate state.
For the first time in this source the ultra-soft component of 
the accretion disk, which black hole binaries display in intermediate or
high-soft states, was clearly detected and measured 
thanks to the high spectral capabilities of XMM-Newton.
\end{abstract}

\section*{Introduction}
The source GRS 1758-258 was discovered in 1990 with the 
SIGMA soft $\gamma$-ray telescope at about 5$^{\circ}$ from the 
Galactic Center \cite{man90}.
The hard spectrum extending up to 200-300 keV \cite{gil93,kuz99},
very similar to the Cyg X-1 spectrum, strongly suggests that this source
is an accreting black hole in a galactic binary system with a low mass 
companion star.
The source was then observed in radio and two symmetrical radio lobes were
detected at 6 cm with the VLA \cite{rod92} on either sides of 
a point-like radio source close to the X-ray source. 
The radio point source position
was compatible with both the SIGMA error circle and the much smaller
Rosat error circle (10$''$ radius) of GRS 1758-258.
In spite of a large drop in hard X-ray flux detected with SIGMA in 1991-1992
and some claims of sporadic appearence of a soft component \cite{mer94},
no spectral transitions have ever been clearly observed from this source.
We present here the first convincing detection of an ultra-soft disk 
emission and of a spectral transition in GRS 1758-258 during a XMM-Newton 
observation.

\section*{XMM-Newton Observations and Results}
The source was observed on 19$^{th}$ September 2000 with XMM-Newton 
for about 10 ks, with EPIC cameras EMOS 1 in timing mode, EMOS 2 in
imaging refresh frame store (RFS) mode and PN in small window mode 
\cite{tur01}. These modes were selected to avoid as much as 
possible pile-up effect, expected for such bright source. The medium 
filter was selected to reduce potential optical loading on the CCDs.
We report here preliminary results from the MOS camera data.
Data reduction was performed using the XMM SAS (Science Analysis Software), 
the standard XSPEC, and the XRONOS packages.
The observation was contaminated by a large flux of ``soft protons'' 
background events; however, thanks to the strength of the source, 
the signal to noise ratio remained very high.

We have used the MOS 2 data to build an image and a spectrum of the source.
Fig. 1 (left) shows the 0.2-10 keV image of the source obtained using 
the central MOS 2 CCD, which was employed in RFS mode for a 
effective integration time of 1325 s.
The count distribution is compatible with a point-like source positioned
at (2000 equinox)
$R.A. ~= ~18^{h} ~01^{m} ~12.5^{s} $ ~ $Dec. ~= ~-25^\circ ~44' ~40''$
with an error radius of 5$''$.

To source spectrum was derived by applying standard cuts to events collected 
within 30$''$ from the source center. Background was estimated using offset 
regions in the same central CCD. The source rate of 11.9 events/frame induced 
a non negligible pile up \cite{bal99}. While no attempt has been made at this 
stage for correcting for it, its effect on the determination of the spectral 
shape was found to be within the statistical error bars of the derived model 
parameters. Its influence on the absolute flux is however much more important, 
and we roughly estimate that it induces a flux loss by a factor of $\sim$ 1.5.

Data were rebinned to reach 20 counts per bin and 
the derived source count spectrum in the range 0.2-10 keV
was compared to several models.
As demonstrated in Fig. 2 (left), a simple power law,
with a reduced chi-square $>$ 3, does not fit the data, 
and the residuals indicate the need to include a soft component. 
The chi-square reaches acceptable values when a soft black body component
is included. 
In Table 1 we report the best fit parameters ($\chi^{2}_{\nu}=1.026$)
for a model of a power-law plus a black body and the unfolded data are 
compared to the model in Fig. 2 (right). 
The soft component with a temperature of $\approx$ 0.3 keV is clearly 
detected, in addition to a power-law with photon index of $\approx$ 2.0,
and reaches a fraction of 15 $\%$ of the total absorbed
0.2-10 keV flux, flux which amounts to 3.7~10$^{-10}$~ph~cm$^{-2}$~s$^{-1}$.
The column density is 1.7 10$^{22}$~cm$^{-2}$ and the derived
source luminosity in the 0.2-10 keV band
at 8 kpc is 8.7~10$^{36}$~erg~s$^{-1}$, out of which $\approx$ 30 $\%$
is due to the black-body component.
No iron lines or other relevant features were detected.

Fig. 3 (left) reports the source light curve with time bins of 1.75 s
obtained using the EMOS 1 data, collected in timing mode for a total 
exposure of 9865 s.
In timing mode the EPIC instruments record the position on
one axis only and the arrival time with a 1.75 ms resolution.
The power density spectrum (PDS) was built using events in the central 
part of the CCD, and grouping the light curve in 21 ms bins. 
The spectrum regrouped in 14 channels after subtraction of the
statistical noise 
is displayed in units of rms$^2$ Hz$^{-1}$ in Fig. 3
(right). It can be modeled by a broken power-law with flat slope
below a break frequency $\nu_{_B}$ of 1.48 Hz, and with slope -1.32 
above $\nu_{_B}$ and
normalization of 7.53 10$^{-3}$ rms$^2$ Hz$^{-1}$ at $\nu_{_B}$.

\section*{Comparison with Previous Results}

In Fig. 1 (right) the XMM-Newton error circle of GRS 1758-258 is
reported on the optical image obtained by Marti et al. \cite{mar98} and 
compared to the Rosat error circle of 10$''$ radius and to the VLA position 
of the point radio source (named VLA C) found at the origin of the extended 
radio lobes.
The XMM error circle intersects the Rosat circle and includes the 
VLA C position confirming the identification of GRS 1758-258 with 
the radio source at the origin of the relativistic jets.

The spectrum obtained with XMM-Newton can be compared to previous X-ray 
spectra
obtained with ASCA \cite{mer97} or RXTE \cite{lin00}. In all cases it is clear
that a spectral evolution took place, since during the XMM observation 
the source clearly displayed a significant soft component accounting for 
more than 15 $\%$ of the measured X-ray flux, component which was never 
clearly detected before.
Compared to the ASCA measurement the 1-10 keV flux also increased by 
a factor $>$ 1.5, and the power-law steepened from 1.7 to 2.0.
Variability characteristics of the source also changed.
In Fig. 3 (right) we report, on the XMM measured PDS, 
the model of PDS derived with RXTE data \cite{smi97}. This plot shows
the Belloni-Hasinger effect typical of black hole systems 
in low-hard state \cite{bel96}. However
the low level of the flat slope and the fact that $\nu_{_B}$ 
reaches values $>$ 1 Hz show that the source state during this 
observation was strikingly close to the intermediate states of Cyg X-1
\cite{bel96}.

We conclude that the source was probably in an intermediate state
considerably different from the standard low-hard state which the source
displayed since its discovery. 
The presence of a soft excess was claimed in the past but not fully 
demonstrated and the disk parameters could not be derived.
A more dramatic state transition was very recently 
observed with RXTE \cite{smi01,smi01b} when the source entered a 
very soft state in March 2001 and in that occasion the black body 
component was detected with a temperature compatible with the values 
reported here.

\begin{table}
%[!h]
%\caption{\textsc{Best-fit Model of P-L + Black-Body }}
\caption{Best-Fit Parameters for a Power Law plus Black Body Model}
\begin{tabular}{ll}
\hline \hline
Parameter & Value (errors at 90$\%$ c.l.)\\
\hline \\
$N_{H} ~(10^{22} \ cm^{-2})$ & $1.74 \pm 0.07$\\
$\alpha_{pl}$ & $1.99 \pm 0.09$\\
$N_{pl} ~(ph ~keV^{-1} cm^{-2} s^{-1}) ~~~~~~~~~~~~~~~~~~~~~~~~~~~$ & $0.13 \pm 0.02$\\
$kT_{bb} ~(keV)$ & $0.32 \pm 0.02$\\
$N_{bb} ~(L_{36}/D^{2}_{10})$ & $3.95  \pm 0.55 $\\
\hline
\end{tabular} 
\end{table}

\section*{Conclusions}
We have presented preliminary results from a XMM-Newton observation 
of the microquasar GRS 1758-258 in September 2000. 
The main findings are the following.
\\
1) The source position is determined with 5$''$ error radius and  
is compatible with the position of the radio point source at the center
of the 2 radio jets.
\\
2) The spectrum of GRS 1758-258 cannot be fit by a simple power-law and a 
soft component (accounting for 15 $\%$ of the measured flux) must be included.
\\
3) The soft component can be described by a black body of kT = 0.3 keV
while the power-law requires a photon spectral index of 2.0.
The total luminosity in 0.2-10 keV band increased by factor $>$ 1.5
with respect to the 1995 ASCA observation.
\\
4) The power density spectrum can be described with the standard flat 
plus power-law function seen in black hole binaries in low-hard state, 
but the rms has decreased and break frequency has increased significantly 
with respect to the values found previously for this source with RXTE.

This observation clearly reveals for the first time in this source 
the ultra-soft spectral signature of an accretion disk,
as seen in many confirmed black hole binary systems. 
Moreover the strength of the disk emission,
the steeper power-law and the different variability characteristics 
indicate that the source underwent a state transition leaving its standard 
low-hard state to enter a typical intermediate state, as observed in 
other black hole binaries like Cyg X-1. 

These preliminary results, though need to be confirmed by deeper and 
more complete analysis, further support the black hole nature of
this source and strengthen the link between relativistic jets and
black holes.
\begin{figure} % fig 1
\centerline{\epsfig{file=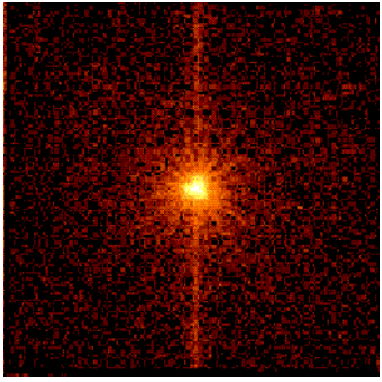,width=5.cm}
 ~~~~~~~~~~~~ \epsfig{file=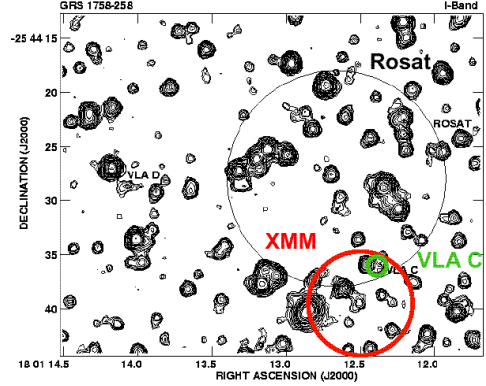,width=6.4cm}
}
\vspace{10pt}
\caption{XMM Newton 0.2-10 keV image of GRS 1758-258 from
the central CCD (11 arcmin size) of EMOS 2 (left). 
XMM-Newton error circle of GRS 1758-258 (5 $''$ radius) reported
on the optical image of the field [8]
and compared to the Rosat error circle of the X-ray source
and to the VLA error circle of the radio point source (named VLA C)
at the origin of the jets (right).}
\label{fig1}
\end{figure}

\begin{figure} % fig 2
\centerline{\epsfig{file=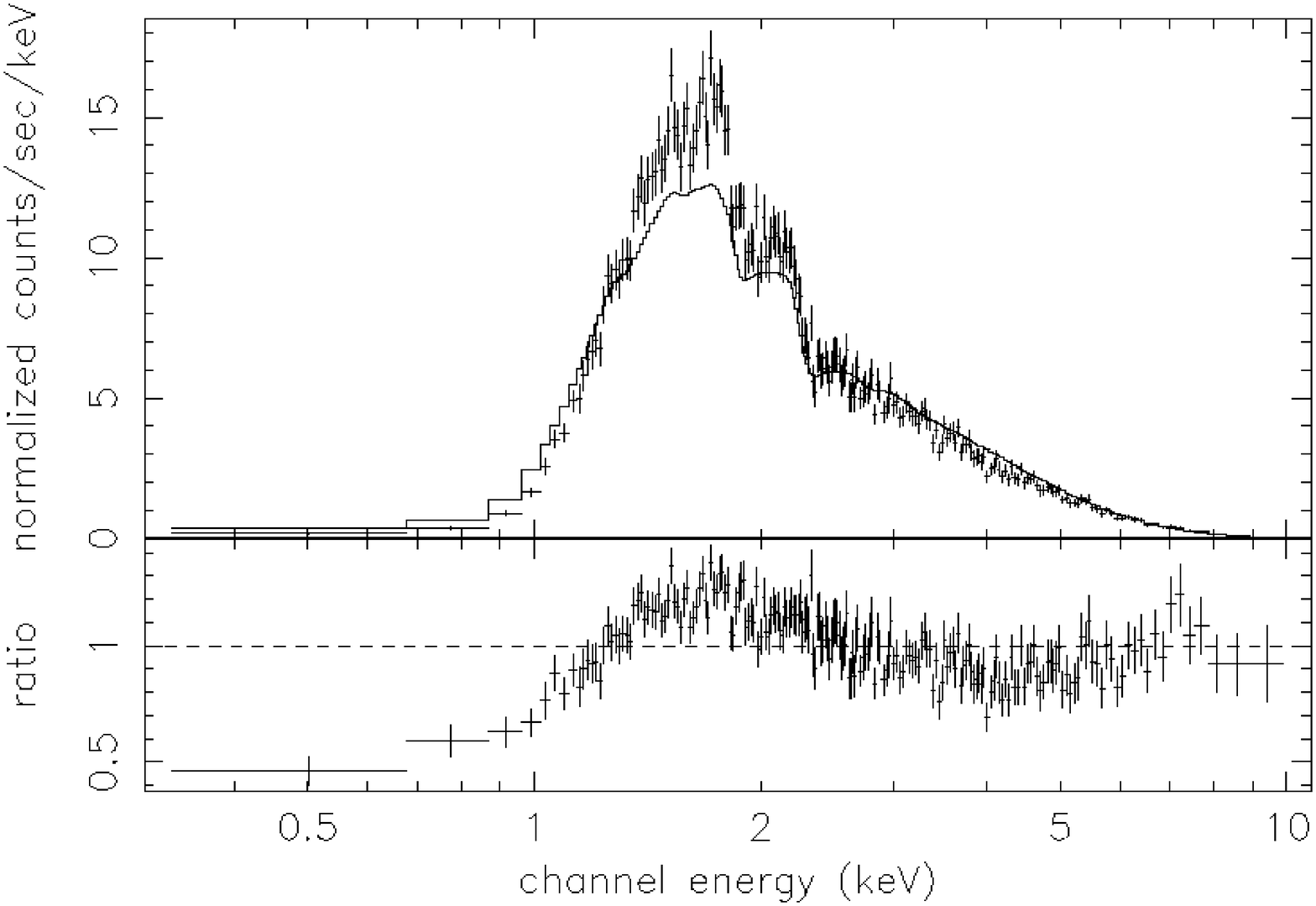,width=7.1 cm}
~~ \epsfig{file=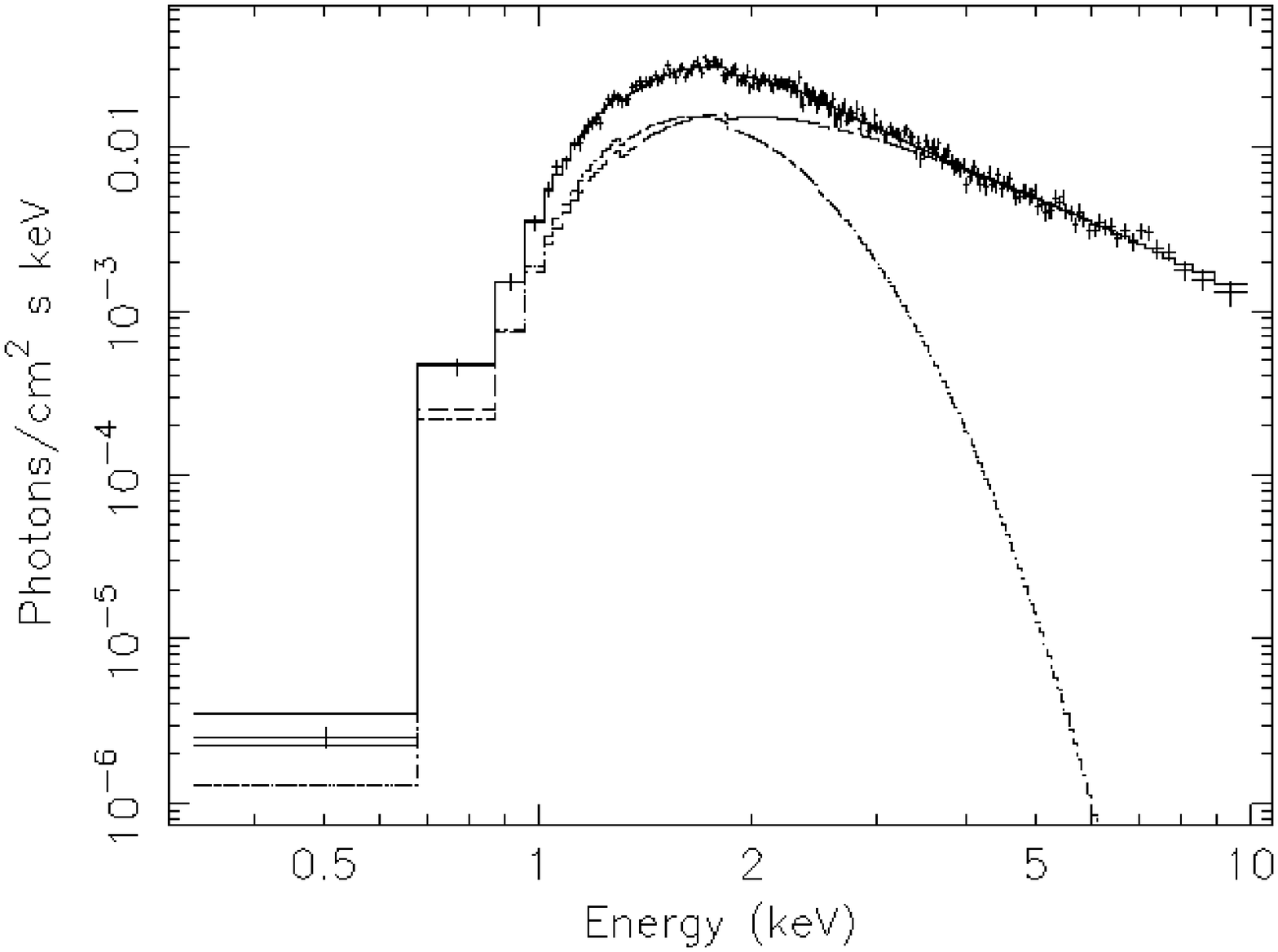,width=6.5 cm} }
\vspace{10pt}
\caption{GRS 1758-258 folded spectrum obtained from EMOS 2 data
compared to the best fit power-law model. Residuals indicate
the need to include a soft component (left).
The unfolded spectrum compared to the best fit model 
of a power-law and a black body (right).}
\label{fig2}
\end{figure}

\begin{figure} % fig 3
\centerline{\epsfig{file=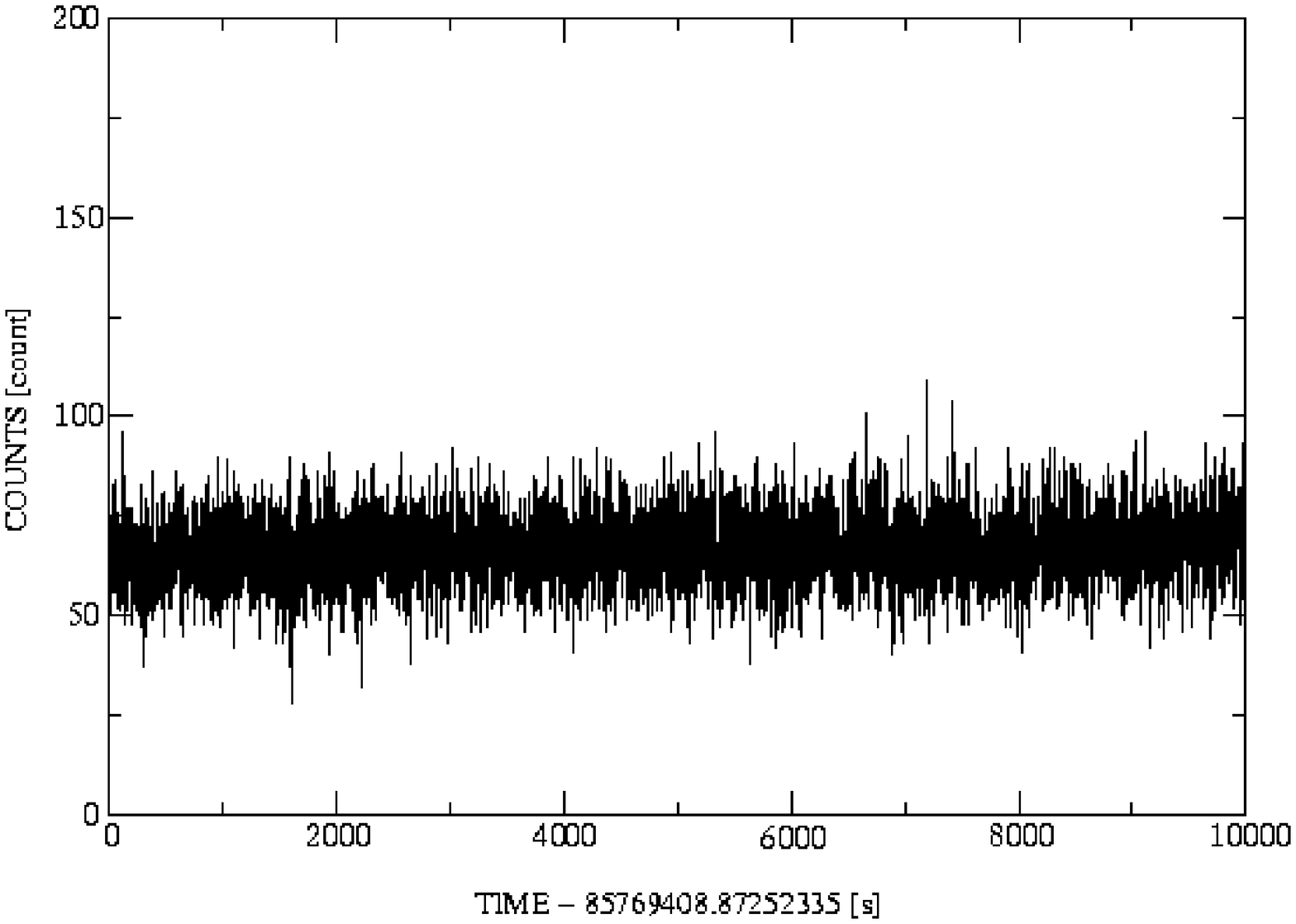,width=6.9cm}
~~ \epsfig{file=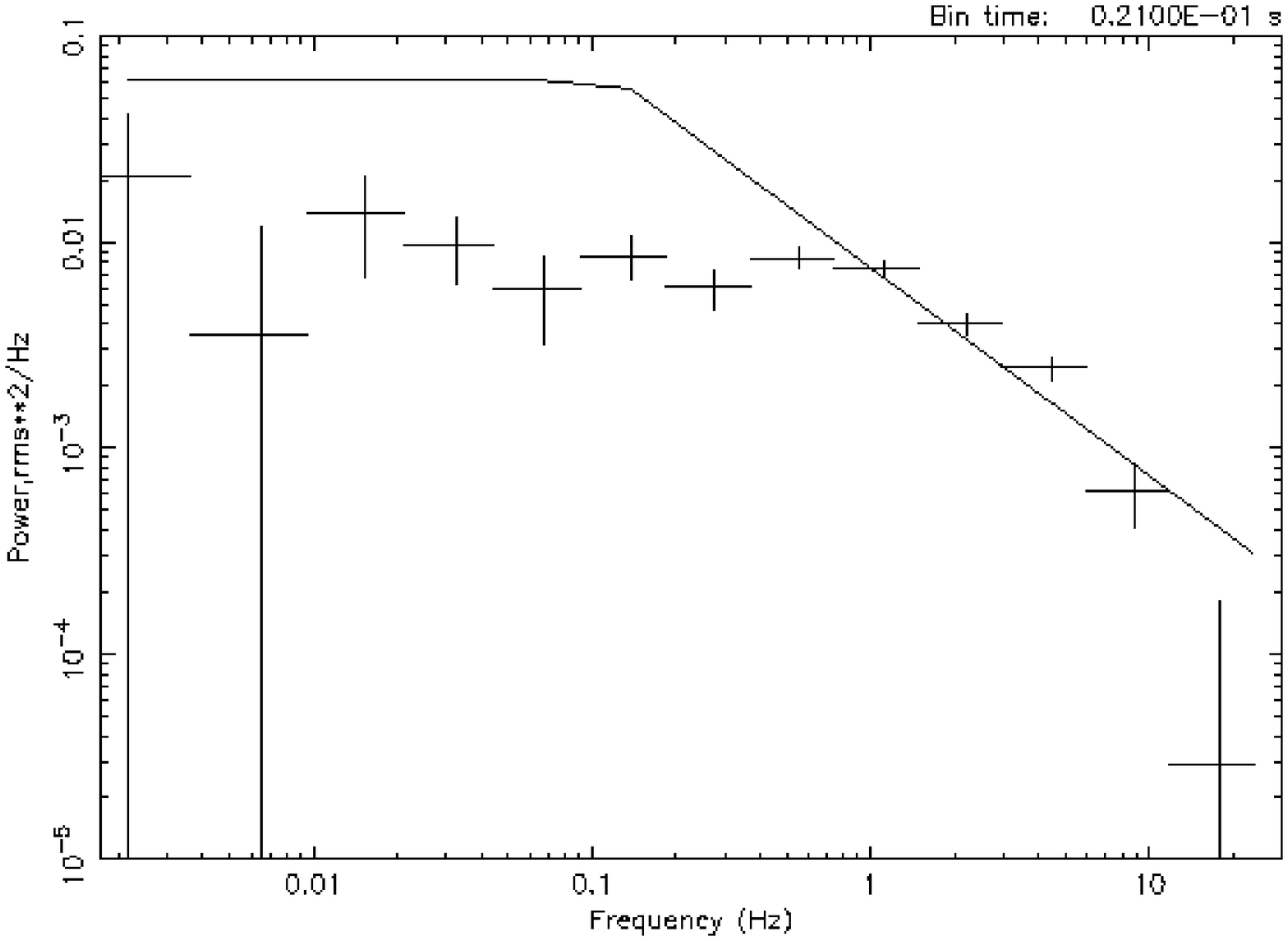,width=7cm} }
\vspace{10pt}
\caption{GRS 1758-258 light curve from the EMOS 1 data in timing mode 
with bins of 21 ms (left).
Power Density Spectrum (PDS) in units of rms$^2$ Hz$^{-1}$ derived 
from the same data and compared to the best fit model of a broken power 
law found in previous observations of the source (left).}
\label{fig3}
\end{figure}


\begin{references}
\bibitem{man90} Mandrou P., {\it IAUC 5032}, (1990).
\bibitem{gil93} Gilfanov M., et al., {\it ApJ}\ {\bf 418}, 844, (1993). 
\bibitem{kuz99} Kuznetsov S. I., et al., {\it Ast.L.}, {\bf 25(6)}, 351, (1999).
\bibitem{rod92} Rodriguez L. F., Mirabel I. F. \& Mart\'\i \  J., {\it ApJ}, {\bf 401}, L15 (1992).
\bibitem{mer94} Mereghetti S., Belloni T., Goldwurm A., {\it ApJ},  {\bf 433}, L21 (1994).
\bibitem{tur01} Turner M.J.L., et al., {\it A$\&$A}\ {\bf 365}, L27 (2001).
\bibitem{bal99} Ballet J., {\it A$\&$A Suppl. Ser.}\ {\bf 135}, 371 (1999).
\bibitem{mar98} Mart\'\i \ J., et al., {\it A$\&$A}\ {\bf 338}, L95 (1998).
\bibitem{mer97} Mereghetti S., et al.,{\it ApJ}, {\bf 476}, 829, (1997).
\bibitem{lin00} Lin D., et al., {\it ApJ}, {\bf 532}, 548, (2000).
\bibitem{smi97} Smith D.M., et al., {\it ApJ}, {\bf 489}, L51, (1997).
\bibitem{bel96} Belloni T., et. al., {\it ApJ}, {\bf 472}, L107, (1996).
\bibitem{smi01} Smith D.M., et al., {\it IAUC 7595}, (2001).
\bibitem{smi01b} Smith D.M., et al., {\it ApJL}, submit. (astro-ph/0103381) 
(2001).

\end{references}
\end{document}